\documentclass[twocolumn,prd,floatfix,showpacs]{revtex4}
\usepackage{epsfig}

\newcommand{\gtrsim}{\,\rlap{\lower3.7pt\hbox{$\mathchar\sim$}}
\raise1pt\hbox{$>$}\,}
\newcommand{\lesssim}{\,\rlap{\lower3.7pt\hbox{$\mathchar\sim$}}
\raise1pt\hbox{$<$}\,}

\begin{document}

\title{Neutrino Dark Energy With More Than One Neutrino Species}

\author{Ole Eggers Bj{\ae}lde}
\email{oeb@phys.au.dk}
\affiliation{Department of Physics and Astronomy, University of Aarhus,
Ny Munkegade, DK-8000 Aarhus C, Denmark}
\affiliation{Center for Cosmology and Particle Physics, Department of Physics, New York University, New York, NY
10003}
\author{Steen Hannestad}
\affiliation{Department of Physics and Astronomy, University of Aarhus,
Ny Munkegade, DK-8000 Aarhus C, Denmark}
\date{{\today}}

\begin{abstract}
The mass varying neutrino scenario is a model that successfully
explains the origin of dark energy while at the same time solves the
coincidence problem. The model is, however, heavily constrained by
its stability towards the formation of neutrino bound states when
the neutrinos become nonrelativistic. We discuss these constraints
and find that natural, adiabatic, stable models with the right
amount of dark energy today do not exist. Secondly, we explain why
using the lightest neutrino, which is still relativistic, as an
explanation for dark energy does not work because of a feedback
mechanism from the heavier neutrinos.

\end{abstract}

\maketitle

\section{Introduction} 
Precision observations of the cosmic microwave background
\cite{Bennett:2003bz,Spergel:2003cb,Spergel:2006hy}, the large scale
structure of galaxies \cite{Tegmark:2006az}, and distant type Ia
supernovae
\cite{Riess:1998cb,Perlmutter:1998np,Astier:2005qq,Wood-Vasey:2007jb}
have led to a new standard model of cosmology in which the energy
density is dominated by dark energy with negative pressure, leading
to an accelerated expansion of the universe.

A very interesting proposal is the so-called mass varying neutrino
(MaVaN) model~\cite{Hung:2000yg,Gu:2003er,Fardon:2003eh} in which a
light scalar field couples to neutrinos, see also
\cite{Peccei:2004sz,Schrempp:2006mk,Ringwald:2006ks,Kaplan:2004dq,Barger:2005mn,Cirelli:2005sg,Barger:2005mh,Gu:2005pq,Li:2004tq,Afshordi:2005ym,Brookfield:2005td,Brookfield:2005bz,Takahashi:2006jt,Spitzer:2006hm,Fardon:2005wc,GonzalezGarcia:2007ib,Bjaelde:2007ki,Ichiki:2007ng,Gu:2007mi,Kaloper:2007gq,Amendola:2007yx,Wetterich:2007kr,Bean:2007nx,Bean:2007ny,Takahashi:2007ru,Gu:2007gy,Bauer:2007gf,Brouzakis:2007aq,Bhatt:2007ah,Anchordoqui:2007iw,Pettorino:2008ez,Mota:2008nj,Bernardini:2008pn,Schrempp:2008zz,Avelino:2008zz,Ichiki:2008rh,Ichiki:2008st,Valiviita:2008iv,Bhatt:2008hr,Li:2008fa}
for background. Because of the coupling, the mass of the scalar field
does not have to be as small as the Hubble scale but can be larger,
while the model still accomplishes late-time acceleration.

In this paper we discuss the different criteria that need to be
fulfilled in order to have a stable, adiabatic mass varying neutrino
model with the correct cosmology today - i.e.
$\Omega_{\mathrm{DE}}=0.7$, $\Omega_{\mathrm{M}}=0.3$ and $w\sim-1$.
Our aim is to show that it is only possible as long as the scalar
field potential resembles a cosmological constant - and hence the
model loses its prediction of the current neutrino mass and fails to
solve the coincidence problem.

In addition, we discuss the suggestion that the lightest neutrino,
which can be relativistic today, may be responsible for dark energy.
We find that there is evidence that the relativistic neutrino
will feel an instability towards the formation of neutrino nuggets.

In the next section we briefly review the formalism needed to study
mass varying neutrinos and in Sec.~\ref{sec:3} we discuss the different
criteria in the MaVaN framework. In Sec.~\ref{sec:4} we discuss MaVaNs with
a relativistic neutrino and in Sec.~\ref{sec:5} we conclude.

\section{Mass Varying Neutrinos} 
The idea in the mass varying neutrino
scenario~\cite{Hung:2000yg,Gu:2003er,Fardon:2003eh} is to introduce
a coupling between neutrinos and a light scalar field and to
identify the coupled fluid with dark energy. In this scenario the
neutrino mass $m_\nu$ is generated from the vacuum expectation value
(VEV) of the scalar field. Thus at scale factor $a$ the pressure $P_\nu(m_\nu(\phi),a)$
and energy density $\rho_\nu(m_\nu(\phi),a)$ of the uniform neutrino
background contribute to the effective potential $V(\phi,a)$ of the
scalar field in the following way

\begin{equation}
 \centering
 V(\phi) = V_\phi(\phi)+(\rho_\nu-3 P_\nu)
 \label{eq:eff}
\end{equation}

where $V_\phi(\phi)$ denotes the fundamental scalar potential.

The energy density and pressure of the scalar field are given by the
usual expressions,

\begin{eqnarray}
 \centering
 \rho_\phi(a)&=&\frac{1}{2a^2}\dot{\phi}^2+V_\phi(\phi),\nonumber\\
P_\phi(a)&=&\frac{1}{2a^2}\dot{\phi}^2-V_\phi(\phi).
\end{eqnarray}

Defining $w=P_{\rm DE}/\rho_{\rm DE}$ to be the equation of state of
the coupled dark energy fluid, where $P_{\rm DE}=P_\nu+P_\phi$
denotes its pressure and $\rho_{\rm DE}=\rho_\nu+\rho_\phi$ its
energy density, the requirement of energy conservation gives,

\begin{equation}
\centering \dot{\rho}_{DE}+3H\rho_{DE}(1+w)=0. \label{eq:Econserv}
\end{equation}

Here $H\equiv \frac{\dot{a}}{a}$ and we use dots to refer to the
derivative with respect to conformal time. Taking
Eq.~(\ref{eq:Econserv}) into account, one arrives at a modified
Klein-Gordon equation describing the evolution of $\phi$,

\begin{equation}
\ddot{\phi}+2H\dot{\phi}+a^2
V_\phi^{\prime}=-a^2\beta(\rho_\nu-3P_\nu). \label{eq:KG}
\end{equation}

Here and in the following primes denote derivatives with respect to
$\phi$ ($^\prime =
\partial/\partial \phi$) and $\beta=\frac{d {\rm log}m_\nu}{d\phi}$
is the coupling between the scalar field and the neutrinos.

In the nonrelativistic case $P_\nu\simeq0$, such that
Eq.~(\ref{eq:eff}) takes the form

\begin{equation}
V=\rho_\nu+V_\phi\label{eq:effNR}
\end{equation}

Assuming the curvature scale of the potential and thus the mass of
the scalar field $m_\phi$ to be much larger than the expansion rate
of the Universe,

\begin{equation}
\label{eq:mphi}
V^{\prime\prime}=\rho_\nu\left(\beta^{\prime}+\beta^2\right)+V_\phi^{\prime\prime}\equiv
m_\phi^2\gg H^2,
\end{equation}

the adiabatic solution to the equation of motion of the scalar field
in Eq.~(\ref{eq:KG}) applies \cite{Fardon:2003eh}. As a consequence,
the scalar field instantaneously tracks the minimum of its effective
potential $V$

\begin{equation}
 \centering
 V^{\prime}=\rho_\nu^{\prime}+V_\phi^{\prime}=\beta\rho_\nu+V_\phi^{\prime}=0
 \label{eq:phi}
\end{equation}

MaVaN models has a possibility of becoming unstable on sub-Hubble
scales $m_\phi^{-1}<a/k<H^{-1}$ in the nonrelativistic regime of
the neutrinos, where the perturbations $\delta\rho_\nu$ evolve
adiabatically.

In Ref.~\cite{Bjaelde:2007ki} it is shown that the equation of
motion for the neutrino density contrast\footnote{For more
information on the evolution of perturbations see
\cite{Bjaelde:2007ki,Bean:2003fb,mb,Hannestad:2005ak,domenico:2002,Koivisto:2005nr,Amendola:2003wa}.}
$\frac{\delta\rho_\nu}{\rho_\nu}$ in the regime
$m_\phi^{-1}<a/k<H^{-1}$ can be written as

\begin{eqnarray}\label{denseom}
\ddot{\delta}_\nu
+H\dot{\delta}_\nu+\left(\frac{\delta
P_\nu}{\delta\rho_\nu}k^2-\frac{3}{2}H^2\Omega_{\nu}\frac{G_{\rm
eff}}{G}\right)\delta_\nu\nonumber\\
=\frac{3}{2}H^2\left[\phantom{\frac{|}{|}}\Omega_{\rm CDM}\delta_{\rm
CDM}+\Omega_{b}\delta_{b}\phantom{\frac{|}{|}}\right]
\end{eqnarray}

where

\begin{eqnarray}
G_{\rm eff}&=&G\left(1+\frac{2\beta^2 M^2_{\rm
pl}}{1+\frac{a^2}{k^2}
\{V_{\phi}^{\prime\prime}+\rho_\nu\beta^{\prime}\}}\right)\,\label{Geff} \\\nonumber \mbox{and}\nonumber\\
\Omega_i&=&\frac{a^2\rho_i}{3H^2M^2_{\rm pl}}.
\end{eqnarray}

Since neutrinos not only interact through gravity, but also through
the force mediated by the scalar field, they feel an effective
Newton's constant $G_{\rm eff}$ as defined in Eq.~(\ref{Geff}). The
force depends upon the MaVaN model specific functions $\beta$ and
$V_{\phi}$ and takes values between $G$ and $G(1+2\beta^2 M^2_{\rm
pl})$ on very large and small length scales, respectively.

In certain cases of strong coupling neutrinos suffer an instability
towards clumping in which case they stop behaving as dark energy
\cite{Afshordi:2005ym}. In Ref.~\cite{Bjaelde:2007ki} a criterion
for the stability was developed. $\left(1+\frac{2\beta^2 M^2_{\rm
pl}}{1+\frac{a^2}{k^2}
\{V_{\phi}^{\prime\prime}+\rho_\nu\beta^{\prime}\}}\right)\Omega_\nu\delta_\nu<\Omega_{\rm
CDM}\delta_{\rm CDM}+\Omega_{b}\delta_{b}$. This can be recast in a
more convenient form $\frac{2\beta^2 M^2_{\rm pl}}{1+\frac{a^2}{k^2}
\{V_{\phi}^{\prime\prime}+\rho_\nu\beta^{\prime}\}}\Omega_\nu<\Omega_{\rm
M}$, where we have neglected the effect of baryons compared to cold
dark matter and we have assumed the density contrasts of roughly the
same order.

\section{Model Requirements}\label{sec:3} 

In summary we have the following different criteria for an adiabatic
MaVaN model.

\begin{itemize}
\item[1)] The model needs to satisfy current observations i.e.
$w\sim-1$. This is easily fulfilled by demanding $\dot\phi\sim0$
since we know the neutrino contribution to both pressure and energy
density is smaller than the scalar field contribution.
\item[2)] We want a neutrino mass at present that satisfies the current
neutrino mass bounds from cosmology, that is
$m_\nu\lesssim1\mathrm{eV}$. Using this, we get a maximal value for
the neutrino density parameter
$\Omega_\nu\lesssim0.02$ \cite{Spergel:2006hy,Tegmark:2006az,Goobar:2006xz}.
\item[3)] The model must produce the right cosmology, i.e.
$\Omega_{DE}\sim0.70$. This gives us roughly
$\frac{\rho_\phi}{\rho_{\nu}}\sim\frac{V_\phi}{\rho_{\nu}}>35$.
\item[4)] In order to have a prediction for the current neutrino mass,
we are looking for an adiabatic model that continuously tracks the
minimum of the effective potential, i.e.
$V^\prime=V_\phi^\prime+\beta\rho_\nu=0$. In addition, to avoid
severe fine-tuning, we demand $m_\phi^{-1}\ll
H^{-1}$ \cite{Kaplinghat:2006jk,Afshordi:2005ym}. Hence
$V_\phi^{\prime\prime}+(\beta^2+\beta^\prime)\rho_\nu\gg H^2$, using
a constant coupling gives us
$V_\phi^{\prime\prime}+\beta^2\rho_\nu\gg H^2$.
\item[5)] We want our model to be stable towards the formation of
neutrino nuggets at present. This gives the requirement
$\frac{2\beta^2 M^2_{\rm pl}}{1+\frac{a^2}{k^2}
\{V_{\phi}^{\prime\prime}+\rho_\nu\beta^{\prime}\}}\Omega_\nu<\Omega_{\rm
M}$. Since we are in the adiabatic regime, we are looking on scales
where
$k/a\sim\sqrt{V_{\phi}^{\prime\prime}+\rho_\nu\beta^{\prime}}$.
Combined with the fact that $\Omega_{\rm M}\sim0.30$, this criterion
simply reduces to $\frac{2\beta^2 M^2_{\rm pl}}{2}=\beta^2 M^2_{\rm
pl}<15$.
\end{itemize}

Combining the criteria above and using
$H^2=\frac{\rho_{\rm{total}}}{M_{\rm{pl}}^2}\sim\frac{V_\phi}{M_{\rm{pl}}^2}$,
we can see that in the case of a constant coupling we get
$\frac{V_\phi^{\prime\prime}}{V_\phi}+\beta^2\frac{\rho_\nu}{V_\phi}\gg
\frac{V_\phi}{M_{\mathrm{PL}}^2V_\phi}$, which reduces to
$\frac{V_\phi^{\prime\prime}M_{\mathrm{PL}}^2}{V_\phi}+\frac{15}{35}\gg
1$ or roughly

\begin{equation}
 \centering
 \frac{V_\phi^{\prime\prime}M_{\mathrm{PL}}^2}{V_\phi}\gg 1.
 \label{eq:cons}
\end{equation}

We propose potentials that resemble those originally proposed in
Ref.~\cite{Fardon:2003eh}, and in order to be able to explain the
coincidence problem we want a potential as simple as possible
without any type of cosmological constant\footnote{Or scalar field
potentials that resemble that of a cosmological constant - e.g. a
very small fractional power-law.}. However all single field
potentials previously studied in this context do not fulfill the
constraint Eq.~(\ref{eq:cons}).

Hence we need a sort of hybrid model as suggested by
\cite{Fardon:2005wc}. In this type of model we have effectively two
minima. One is the true minimum which has zero energy density and
the other is the false minimum in which our current universe sits.
The offset between the two minima is then what is referred to as the
dark energy density. The dynamics of the scalar field is given by
the movement of the false minimum with the dilution of the universe.
One very nice feature of this model is that it is implemented in
supersymmetry and hence explains the stability of a very small scalar field
mass. However, since we have an offset it can be interpreted as an
effective cosmological constant and as such the hybrid model fails
to address the coincidence problem. In addition, to keep the model
stable it is suggested that the lightest neutrino is still
relativistic and that we associate dark energy with this. We address
this issue below.

\section{MaVaN model with a relativistic neutrino}\label{sec:4} 

We aim at keeping the model as simple as possible, so we propose the
existence of just one scalar field with a democratic
coupling to all light neutrino species \footnote{One could of course
argue that a model with three scalar fields each coupled to one neutrino
state would be a natural model to investigate, however, we are merely
interested in a proof of concept here, so we stick to the simpler
scenario with just one scalar field. In addition this is the standard 
assumption in the literature.}. In that case it is assumed
that the lightest neutrino does not clump because of pressure, while
the two heavier neutrinos become unstable. In the following we will
argue that this is not possible since a feedback from the growth of
the heavy neutrino perturbations will cause the relativistic
neutrino perturbation to grow as well. The problem in this scenario
can be neatly illustrated by the following
equation\cite{Brookfield:2005td}

\begin{equation}
\centering
 \delta\rho_\nu=\frac{1}{a^4}\int q^2dqd\Omega\epsilon
 f_0(q)\Psi+\delta\phi\beta(\rho_\nu-3P_\nu).
 \label{eq:neutrinop}
\end{equation}

The equation explains the growth of the neutrino perturbation and
applies to both the interaction between a relativistic neutrino and
a scalar field as well as that of a nonrelativistic neutrino and a
scalar field.

Let us consider a system consisting of three neutrinos, two heavy
nonrelativistic neutrinos (H) and one light relativistic neutrino (L)
both interacting with a scalar field. We only have one scalar field
so the nonrelativistic neutrinos will inevitably become unstable
towards clumping and $\delta\rho_\nu(H)$ grows to very large values.
For an interaction with a nonrelativistic neutrino, the average
scalar field perturbation can be written as \cite{Bjaelde:2007ki}

\begin{equation}\label{averagedphi}
\delta\bar{\phi}=-\frac{\beta\rho_\nu\delta_\nu}{(V_{\phi}^{\prime\prime}+\rho_\nu\beta^{\prime})+\frac{k^2}{a^2}}.
\end{equation}

As we can see the scalar field perturbation is effectively
proportional to the neutrino perturbation. Note that in reality
there will be extra terms in Eq.~(\ref{averagedphi}) since we also
have a relativistic neutrino, however, these terms will be
sub-dominant compared to the terms from the heavier neutrinos. So
$\delta\bar{\phi}$ will grow as a result.

Turning our attention to $\delta\rho_\nu(L)$, we consider the second
term in Eq.~(\ref{eq:neutrinop}). $\delta\bar{\phi}$ is growing
uninhibited to large values, the coupling $\beta$ is for simplicity assumed of
order unity (Planck units)\footnote{Note that this assumption makes no
difference qualitatively, a smaller coupling would only act to delay
the instabilities slightly.}. $(\rho_\nu-3P_\nu)$ is a suppression
factor of the order $m/E$ - this factor will act to delay the growth
of $\delta\rho_\nu(L)$. However, since $\delta\bar{\phi}$ will
continue its growth, the inevitable conclusion is that
$\delta\rho_\nu(L)$ will eventually start to grow. Hence, there
exists a type of feedback mechanism between the neutrinos. Note
that in this treatment we only considered the interaction between
neutrinos of the same type via exchange of the scalar field. If we
include couplings between different neutrinos, the feedback
mechanism will be even stronger. One could of course argue that we
are exactly living in a transition regime when $\delta\rho_\nu(L)$
has still not turned unstable. However, that would require serious
fine-tuning.

We also investigate a MaVaN model with a relativistic neutrino numerically
using the publicly available CMBFAST code \cite{CMBFAST}, where we implement
the scenario with two heavy neutrinos with assumed masses $m_\nu(H)=0.312$ eV and
one species of relativistic neutrinos with assumed mass $m_\nu(L)=0.0001$ eV.
For the scalar field potential we use a Coleman-Weinberg\cite{Coleman:1973jx}
type scalar field potential similar to the one presented in Refs.~\cite{Fardon:2003eh,Bjaelde:2007ki}

\begin{equation}
 \centering
 V_\phi=V_0\,\log{\left(1+\kappa\phi\right)},
\end{equation}

where $\kappa$ and $V_0$ are dimensionful constants, which are fixed by demanding the appropriate amount
of dark energy in the universe today as well as solving for the minimum of the effective potential.
For the mass term of the three neutrinos
we choose the same mass term as in \cite{Bjaelde:2007ki}

\begin{equation}
 \centering
 m_\nu=\frac{m_0}{\phi},
\end{equation}

which results in a coupling $\beta=-\frac{1}{\phi}$\footnote{This definition has the advantage that the value
of the coupling is determined by the VEV of the field itself, which means that there is no extra free parameter involved.}.

The code calculates the background energy density and pressure of different species from standard
integration methods while at the same time it solves for the minimum of the effective potential of
the neutrino-scalar field fluid which is

\begin{equation}
\centering
 V(\phi) = V_\phi(\phi)+(\rho_\nu(H)-3 P_\nu(H))+(\rho_\nu(L)-3 P_\nu(L)),
\end{equation}

where we remember that there are now two heavy (H) neutrinos and one light (L) neutrino. The expression above
can be recast in a more convenient form as \cite{Fardon:2005wc}

\begin{equation}
 \centering
 V(\phi) = V_\phi(\phi)+m_\nu(H) n_\nu(H)+\frac{m_\nu(L)^2 T_\nu^2}{12},
\end{equation}

where we introduced the neutrino number density $n_\nu$ and neutrino temperature $T_\nu$. At the same
time as tracking the background we calculate the perturbations to first order such as

\begin{equation}
 \centering
 \rho_\nu=\rho_\nu+\delta\rho_\nu.
\end{equation}

\begin{figure}[!htb]
\begin{center}
\includegraphics*[width=7cm]{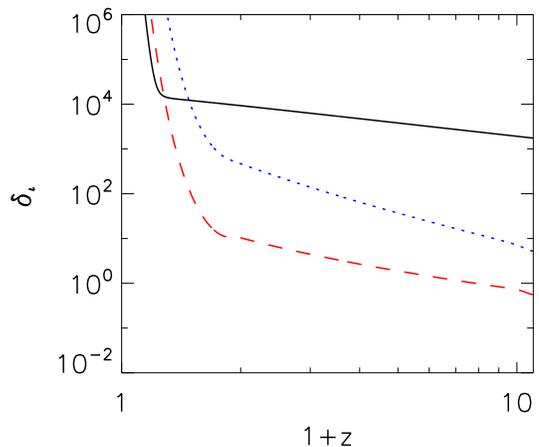}
\caption[]{Density contrasts plotted as a function of redshift for a
system consisting of one light and two heavy MaVaN neutrinos each
interacting with the same scalar field. The scale is $k=0.1\,{\rm
Mpc}^{-1}$ and we choose the current neutrino masses
$m_\nu(L)=0.0001$\,eV and $m_\nu(H)=0.312$\,eV (Note that the choice
of current neutrino masses does not affect the result
qualitatively). We have chosen $\kappa=1\times10^{18}$ in CMBFAST units of
$(M_{\rm PL}/{\rm Mpc})^{-1/2}$ and $V_0=1.11\times10^{-9}\,\,(M_{\rm PL}/{\rm Mpc})^{2}$
to fix the cosmology.
The solid black line is CDM-density contrast defined as $\frac{\delta\rho_{\rm CDM}}{\rho_{\rm CDM}}$,
the dotted blue line is the heavy neutrino density contrast $\frac{\delta\rho_\nu(H)}{\rho_\nu}$,
and the dashed red line is the light neutrino density contrast $\frac{\delta\rho_\nu(L)}{\rho_\nu}$.
The heavy neutrinos grow moderately until the coupling becomes large enough for the
instabilities to set in. The light neutrino is still relativistic,
and its density contrast oscillates as acoustic waves while growing slightly. However, due
to a feedback mechanism, the relativistic neutrino density contrast
tracks that of the nonrelativistic neutrino especially as the growth of the heavy neutrino perturbation becomes unstable; i.e.
both neutrino species will clump. Note that the CDM perturbations
also blow up at late times. This is an effect of the system of
differential equations breaking down as all parameters go to
infinity.} \label{fig:dens}
\end{center}
\end{figure}

This is done for both heavy and light neutrinos using Eq.~(\ref{eq:neutrinop}).
The results are shown in Fig.~\ref{fig:dens}, where we plot the evolution of the
neutrino perturbation and CDM perturbation from a redshift of $z+1=11$ until today.

What happens in Fig.~\ref{fig:dens} is that the perturbations of the heavy neutrinos,
which have already turned nonrelativistic before the range of this plot, are already growing
albeit at a slow pace because of the growth-slowing effect of CDM - see Eq.~(\ref{denseom}). The perturbation of the light
neutrino is oscillating (only visible if zoomed in) in this epoch, but also growing slightly as
a result of the growth of the coupling $\beta$. The growth-slowing effect of CDM is, however,
only temporary and around a reshift of $z+1\sim2$ the heavy neutrino perturbation can no longer
resist the temptation towards unstable growth. The heavy neutrino perturbation explodes and because
of the increased value of $\beta$, the light neutrino perturbation starts exploding as well. We do see
that eventually the CDM perturbation also starts exploding and at this moment the linear code is
no longer valid. However at this moment both neutrino perturbations are already in the nonlinear regime and
will have started forming neutrino bound states. In this regime the MaVaN fluid will no longer act as dark energy
and the model breaks down. In conclusion, as a result of a feedback mechanism, the
blowing behaviour of the nonrelativistic neutrino density contrast
causes the relativistic neutrinos to start clumping as well. Hence the neutrino scalar-field fluid will start acting as a cold
dark matter component (clustering neutrinos) and hence cannot be
attributed to dark energy.

The reason that the relativistic neutrino is able to clump is that
it will acquire an effective mass, thus it cannot be regarded as a
relativistic particle. Unfortunately, we cannot use conventional
bounds to constrain this effect since the evolution of the neutrino
perturbations become nonlinear, i.e. the whole system of equations
we are solving, starting with the modified
Klein-Gordon equation breaks down. This has the effect that all current bounds
are no longer valid, as these are established in the linear regime.

\section{Conclusion}\label{sec:5} 
Single scalar field models can be used to explain late-time
acceleration in the MaVaN scenario. However, in general using these
potentials leads to instabilities towards neutrino bound states
unless certain criteria are relaxed. In order to obtain a stable or
metastable model we direct the reader to Ref.~\cite{Bjaelde:2007ki}
in which one such model is presented. The crucial thing is that the
scalar field potential resembles that of a cosmological constant. In
this case we simply move the fine-tuning problem to that of
explaining the small fraction in the power-law - which unfortunately
is also rather unnatural.

Accordingly it has been suggested to include an extra scalar field
in the treatment. This has some very nice features and is easily
capable of obtaining late-time acceleration as well as
$\Omega_{\mathrm{DE}}=0.7$ today. However, one drawback is the need
for the lightest neutrino to be relativistic today. As was explained
above the feedback mechanism will eventually cause the perturbations
of the relativistic neutrino to start growing. As a result, the MaVaN
fluid will cease to act as dark energy.

\section*{Acknowledgments} 
We would like to thank Lily Schrempp, Carsten van de Bruck, Anthony
Brookfield, David Mota, and Domenico Tocchini-Valentini for their
great collaboration in the paper establishing the underlying
framework for this. In addition OEB would like to thank Neal Weiner
and Subinoy Das for some fruitful discussions and NYU for
hospitality. We also acknowledge the use of the publicly available
CMBFAST code \cite{CMBFAST}.
\section*{References} 

\end{document}